\documentclass[a4paper, 11pt]{article}

\usepackage{epsfig,amssymb,euscript,xspace,color}
\usepackage{amsmath,jheppub_0,mathtools}
\def\bea{\begin{eqnarray}}
\def\eea{\end{eqnarray}}
\def\be{\begin{equation}}
\def\ee{\end{equation}}

\newcommand{\bra}{\langle}
\newcommand{\ket}{\rangle}
\newcommand{\rpl}{r_{\!_+}}
\newcommand{\rmi}{r_{\!_-}}
\newcommand{\Ll}{\mathcal{L}}


\def\cL{{\cal L}}

\def\cL{{\cal L}}

\def\S{{\cal S}}

\begin{document}

\title{Jackiw-Teitelboim gravity and near-extremal BTZ thermodynamics}
\author{ Rohan R. Poojary$^a$}
\affiliation{$^a$Institute for Theoretical Physics, TU Wien,\\
Wiedner Hauptstrasse 8-10, 1040 Vienna, Austria.  }
\emailAdd{rpglaznos@gmail.com}

\abstract{We study the near horizon 2d gravity theory which captures the near extremal thermodynamics of BTZ with excess mass $\delta M$ and excess angular momentum $\delta J =\Ll\, \delta M$ over extremality. We find that the Jackiw-Teitelboim theory is able to capture such departures from extremality with an additional parameter $\Ll$ relating it to the near extremal BTZ configuration. We are able to show this by obtaining new simultaneous near horizon near extremal limits of the BTZ geometry parametrized by $\Ll$. The resulting Jackiw-Teitelboim theory captures the near extremal thermodynamics of such BTZ geometries provided we identify the temperature $T^{(2)}_H$ of the $AdS_2$ geometry in the Jackiw-Teitelboim theory to be $T^{(2)}_H=T_H/(1-\mu\,\Ll)$ where $T_H$ is the BTZ temperature and $\mu$ its chemical potential with $\mu\,\Ll<1$. }

\maketitle



\section{Introduction}
Black holes have always been a fertile area of research with regards to the quantum theory of gravity. The holographic principle motivated by string theory have made possible advances in our understanding of how black holes scramble in fallen information. A quantum information theoretic approach for large $N$ field theories had allowed Hayden, Preskill and others \cite{Hayden:2007cs,Sekino:2008he} to estimate the fastest time scales possible to scramble any local\footnote{Capturing $\mathcal{O}(1)$ degrees of freedom. } perturbation among its microstates to be $t_*\sim \log\mathcal{S}$, $\mathcal{S}$ being the entropy of the system. These works also conjectured that black holes are amongst the fastest scramblers of in fallen information in nature. A precise measure of such a chaotic behaviour was fist performed by checking the rate of scrambling due to a $\mathcal{O}(G_N)$ perturbation of the finely tuned entanglement between the 2 CFTs corresponding to the Thermo-field Double(TFD) state dual to a Schwarzchild black hole in $AdS_3$ \cite{Shenker:2013pqa}. It was also shown that the 4pt \emph{out of time ordered correlators} (OTOCs) which are a good measure of chaotic phenomena, see an exponential growth in their first sub-leading term in $G_N$ at large time separations
\bea
&&1-\frac{\bra W(0)V(t)W(0)V(t)\ket_{\rm otoc}}{\bra WW\ket\bra VV\ket}\sim \tfrac{G_N}{l}e^{\lambda_L t}\cr&&\cr
&{\rm where}&\hspace{1.5cm}\lambda_L=2\pi T_H
\label{lambda_L_schwarz}
\eea     
$T_H$ being the black hole temperature. Arguments of boundedness and analyticity of the 4pt correlators in the complex time plane allowed Maldacena, Shenker \& Stanford to bound the instantaneous value of $\lambda_L$ to be $\leq 2\pi T_H$  in a thermal system \cite{Maldacena:2015waa}. These arguments have since been further utilised to bound subsequent terms in a small parameter expansion of the 4pt OTOC \cite{Kundu:2021mex,Kundu:2021qcx}.
\\\\
However the case of rotating black holes is a bit subtle. 
It was shown that the 4pt OTOC of light operators in a state dual to rotating BTZ exhibits 2 Lyapunov indices $2\pi/\beta_\pm$, each corresponding to the left and right temperatures of the CFT$_2$ \cite{Stikonas:2018ane,Poojary:2018esz,Jahnke:2019gxr}. 
The generalization of the arguments in \cite{Maldacena:2015waa} to thermal system with a chemical potential $\mu$ for space time symmetries revealed that the bound on chaos is modified to be \cite{Halder:2019ric}
\be
\lambda_L\leq\frac{2\pi}{\beta(1-\mu/\mu_c)}
\label{Halder_bound}
\ee 
where $\mu_c$ is the critical value for the chemical potential\footnote{Here critical value implies the maximal value attainable by $\mu$, in case of holographic systems this would correspond to extremal black hole geometries in the bulk.}. The $rhs$ above does translate to the greater of the 2 CFT$_2$ temperatures thus conforming to the lore that black holes are  amongst the fastest scramblers of information. 
It is important to note that there does not yet exist an SYK-like model which can be thought of as an equivalent dual as rotation is absent in 1d. 
It was further shown by taking the periodicity of the spatial boundary coordinate of the BTZ geometry into account that the average $\lambda_L$ is still given by the $2\pi T_H$, whereas the growth of the OTOC leading upto the scrambling time scale shows a sawtooth like behaviour with the $\lambda_{L}^{\rm max}$ being greater of the 2 temperatures of the dual CFT$_2$. 
The disruption of the Mutual Information between the 2 CFT$_2$s representing the TFD dual of the BTZ can also serve as a diagnostic of chaotic behaviour. 
Indeed the Mutual Information between large enough subsystems can provide an upper bound on the behaviour of local correlators in a CFT \cite{Wolf:2007tdq}.
It was using such a measure that chaotic behaviour of black holes was first investigated \cite{Shenker:2013pqa}. 
This analysis crucially revealed that the disruption of the Mutual Information due to an in-falling perturbation is basically governed by the blueshift suffered by the perturbation at late times as it falls into the black hole. 
The same analysis was also repeated for rotating BTZ \cite{Reynolds:2016pmi} with the same conclusions for $\lambda_L$. 
However when similar computation was done for perturbation due to rotating shockwaves in a rotating BTZ it revealed that in general $\lambda_L>2\pi T_H$ and it is bounded by Max$[2\pi/\beta_\pm]$. Further, in certain cases the scrambling time is also governed by such a $\lambda_L>2\pi T_H$ \cite{Malvimat:2021itk}. 
This analysis has been recently generalized to Kerr AdS$_4$ where one analyses the late time disruption of Mutual Information due to rotating shockwaves along the equator \cite{Malvimat:2022oue}. 
Here the shockwaves and subsystems in the 2 boundary CFT$_3$s  corresponding to the dual Kerr geometry are chosen  to respect the axi-symmetry of the geometry. In this case the late time scrambling is indeed seen to be governed by 
\be
\lambda_L=\frac{2\pi}{\beta(1-\mu\,\Ll)}
\label{lambda_L_rot}
\ee     
where $\mu$ is the horizon velocity $w.r.t.$ a stationary boundary observer and $\Ll$ is the shockwave's angular momentum per unit energy. We also find that the $\Ll$ is restricted by turning point analysis to be less than $\mu^{-1}$ for non extremal geometries and approaches $\mu^{-1}$ as the geometry approaches extremality\footnote{In such a case the blueshift reaches a finite limit $c.f.$ eq(3.14) in \cite{Malvimat:2022oue}.}.
\\\\
The analysis of the now famous 1d Sachdev-Ye-Kitaev (SYK) model by Maldacena \& Stanford \cite{Maldacena:2016hyu} showed that the strongly coupled fermions in such  a theory exhibited chaotic behaviour of the form \eqref{lambda_L_schwarz} close to its IR fixed point which in turn exhibits a spontaneously broken 1d conformal symmetry. The chaotic behaviour can be inferred due to the Schwarzian action for time reparametrizations due to this broken conformal symmetry close to the IR.
Following the analysis of Almheiri and Polchinski \cite{Almheiri:2014cka} it was famously shown by Jensen \cite{Jensen:2016pah} and by Maldacena, Stanford \& Yang \cite{Maldacena:2016upp} that the $AdS_2$ dynamics in a specific 2d dilaton model of gravity which captures the near horizon near extremal dynamics of higher dimensional black holes reproduces the expected chaotic behaviour. This behaviour is governed by a  Schwarzian action for deformations of the near horizon $AdS_2$.
The  2d gravitational action was earlier analysed by Jackiw and Teiltelboim and is known as the JT action \cite{Teitelboim:1983ux,Jackiw:1984je}.
Here they showed that the scrambling of local 2d probe fields in the nearly $AdS_2$ bulk has an exponential growth with $\lambda_L=2\pi T_H$ with $T_H$ being the infinitesimally small temperature associated with the $AdS_2$ geometry which was expected to be the near horizon throats of near extremal black holes. 
The analogy between SYK-like models and 2d dilaton theories of gravity like the JT theory has been further generalized to the case of complex SYK model which in certain regimes has been shown to be equivalent to the $\widehat{\rm CGHS}$ model \cite{Afshar:2019axx}. 
The JT action has since been proved to reproduce the near extremal thermodynamics of higher dimensional black holes, for example in BTZ \cite{Almheiri:2016fws,Banerjee:2019vff}, Riessner-Nordstrom black holes in 4d and 5d \cite{Nayak:2018qej}, Kerr black holes in 4d and 5d \cite{Moitra:2019bub}. 
In each of the above cases the JT description of the near extremal BHs  describes the near horizon $AdS_2$ as having the same temperature as the black hole and any analysis of interaction with probe matter fields in this setting reproduces the exponential scrambling behaviour with $\lambda_L=2\pi T_H$. 
\\\\
This contrasts starkly with the above mentioned picture for rotating geometries. One may ask if the  sawtooth like late time behaviour of 4pt OTOC \cite{Mezei:2019dfv,Craps:2020ahu,Craps:2021bmz} which exhibits the instantaneous $\lambda_L^{inst}>2\pi T_H$ can be seen holographically in the IR $i.e.$ in the near horizon analysis of near extremal BTZ. Given that the late time scrambling of mutual information in BTZ  \cite{Malvimat:2021itk}  and Kerr $AdS_4$ \cite{Malvimat:2022oue} is indeed governed by a $\lambda_L>2\pi T_H$, it strongly indicates that there must be a near horizon picture similar to the JT analysis explored till now in literature which supports such a behaviour of scrambling.     
A crucial point to be noted in all these analyses is that the JT action in literature describes the thermodynamics of near horizon fluctuations where all the charges except the mass of the higher dimensional black hole is held fixed. In other words the JT action captures the thermodynamics of configurations with excess mass and entropy over extremality for near extremal BHs. 
\\\\
In this paper we strive to get a better near horizon understanding of the thermodynamics of near extremal black holes in $AdS_3$ when both mass and angular momenta are allowed to change over extremality. In particular we consider extremal BTZ black holes which are made near extremal with the addition of mass $\Delta M$ and angular momentum $\Delta J=\Ll\,\Delta M$ where $0\leq\Ll<1$. We first describe in section 2 how the JT gravity action results from the Einstein-Hilbert action in $AdS_3$ about a near extremal solution after dimensional reduction\footnote{The readers are referred to \cite{Astorino:2002bj} for a 2d supergravity analysis of JT theory.}. We do this with an eye towards describing the thermodynamics of near extremal BTZ with $\Ll=0$. In section 3 we write down new near horizon near extremal limits by considering going near horizon along in-falling null rotating geodesics with angular momentum per unit energy $\Ll$ and show that the $\Ll=0$ case reproduces such limits found in literature \cite{Gupta:2008ki,Sen:2008yk}. The physical motivation for such limits for $\Ll\geq 0$ is as follows: any null perturbation to the the black hole emanating from the $AdS_3$ boundary would follow a null geodesic characterized by its angular momenta. When it reaches the near horizon region along such a geodesic the back reaction to the geometry ought to be described in terms locally smooth coordinates which do not see any coordinate singularity at the horizon. This also determines the exponent of the blueshift \eqref{lambda_L_rot} suffered by such a null perturbation \cite{Malvimat:2021itk,Malvimat:2022oue}. 
This gives a precise relation between the near horizon coordinates, $\Ll$ and the Boyer-Lindquist coordinates for the near extremal BTZ. 
This allows us to show that the equivalent JT description -which we call JT$_\Ll$ for disambiguation, about such a near horizon solution describes near extremal black holes where $\delta J-\Ll\,\delta M=0$. We then proceed to show in section 4 that the near horizon thermodynamics of the JT$_\Ll$ model captures the correct excess mass and entropy above extremality if the near horizon $AdS_2$ black hole has a temperature $T_H^{(2)}$ which is related to the near extremal BTZ temperature $T_H$ as
\be
T^{(2)}_H=\frac{T_H}{(1-\mu\,\Ll)}
\label{T2_to_T_H}
\ee  
where $\mu$ is the chemical potential close to extremality which can be taken to be $1$. We also see that the first law of black hole thermodynamics is reproduced from the thermodynamics of JT$_\Ll$. This constitutes the main result of this paper. 
In section 5 we find repeat  the analysis of Jensen's \cite{Jensen:2016pah} and of Maldacena, Stanford \& Yang \cite{Maldacena:2016upp} for the JT$_\Ll$ theory to show that the 2d probe scalar fields in the JT$_\Ll$ theory scramble information at an exponential rate determined by $T^{(2)}_H$. We wrap up with conclusions and discuss some questions thrown open by the analysis in this paper. 
\section{BTZ and JT gravity}
Extremal BHs have a near horizon metric with an AdS$_2$ factor \cite{Kunduri:2007vf,Figueras:2008qh}. The near extremal fluctuations of the metric can be captured in the JT model after having dim. reduced the rest of the (angular) coordinates. 
\\\\
Let's see this in the case of gravity in AdS$_3$. 
An extensive analysis for the case of BTZ can be found in \cite{Ghosh:2019rcj}. The Einstein-Hilbert action with negative cosmological constant in 3 dimensions is
\bea
I=-\frac{1}{16\pi G_N}\int d^3x \sqrt{g_3}(R_3-2\Lambda)-\frac{1}{8\pi G_N}\int_{\partial}d^2x\sqrt{h}\left(K_3-\frac{1}{l}\right)
\label{AdS_3_action}
\eea
where $\ell$ is the $AdS$ radius in 3d and $G_N$ being the 3d Newton's constant. 
The rotating BTZ solution is generically written as 
\bea
\frac{ds^2}{l^2}&=&\frac{\rho^2d\rho^2}{(\rho^2-r_+^2)(\rho^2-r_-^2)}-\frac{(\rho^2-r_+^2)(\rho^2-r_-^2)dt^2}{\rho^2}+r^2\left(d\phi-\frac{r_+r_-}{\rho^2}dt\right)^2\cr&&\cr
&&\hspace{-1.2cm}{\rm having}\hspace{0.5cm}8G_NM=r_+^2+r_-^2,\,\,8G_NJ=2lr_+r_-.
\label{BTZ_metric_wiki}
\eea
We begin by dim. reducing the action \eqref{AdS_3_action} over the  following class of geometries
\begin{equation}
ds^2=\Phi^{2\alpha}(ds_{(2)}^2)+\Phi^2(dy+A_t dt)^2,
\label{dim_red_metric}
\end{equation} 
yielding \footnote{$\alpha$ can be chosen to be $0$ for 3d.}
\be
\int d^3x\sqrt{g_3}( R_3-2\Lambda)=2\pi \int d^2x\sqrt{g_2}\,\Phi\left(R_2-2\Lambda+\frac{1}{4}\Phi^2 F^2\right),\hspace{0.5cm}{\rm for}\,\,\,\alpha=0
\label{AO_0}
\ee
where the factor of $2\pi$ comes from dimensionally reducing over the periodic coordinate.
We proceed by solving for the field strength as it is completely determined in 2d (upto the charge which is the angular momentum in the 3d geometry).
\be
\nabla_\mu (\Phi^3 F^{\mu\nu})=0  \implies F_{rt}=\sqrt{g_2}\frac{Q}{\Phi^3}
\label{F_eom}
\ee
 Here we have assumed that the geometries we would be concerned with are stationary\footnote{This not only applies to the near extremal near horizon geometries we would be concerned with but also crucially the fluctuations about such a geometry.}.
The action reduces to 
\be
I_{(AO)}=-\frac{1}{8G_N}\int \sqrt{g_2}\Phi\left(R_2-2\Lambda-\frac{Q^2}{2\Phi^4}\right)-\frac{1}{4G_N}\int_\partial \sqrt{h_2}\Phi K_2. 
\label{AO_1}
\ee
One may choose to work with either \eqref{AO_0} wherein the charge $Q=const$ is implied by $e.o.m.$ and stationarity or with \eqref{AO_1} where $Q$ is fixed to be a specific $const.$ For a complete treatment of integrating the gauge field out one can refer to \cite{Ghosh:2019rcj}. 
The above action has additional boundary counter terms which makes the action finite on-shell. 
To obtain the JT theory we expand \eqref{AO_1} about the near horizon value of the dilaton $\Phi$ read off from the near horizon form of the extremal metric
\be
ds^2_{NH}=\frac{ds^2}{\ell^2}=\frac{1}{4}\left[\frac{dr^2}{r^2}-r^2d\tau^2\right]+r_+^2\left(dy+\frac{r}{2r_+}d\tau\right)^2
\label{BTZ_nhext}
\ee 
to linear order $i.e.$ $\Phi=r_+ + \phi$. This yields
\be
I_{\rm JT}=I_{(0)} -\frac{1}{8G_N}\int \sqrt{g_2}\,\phi\left(R_2-2\Lambda+\frac{3Q^2}{2r_+^4}\right)-\frac{\phi_\partial}{4G_N}\int_\partial\sqrt{h_2}\,\, 
\left(
K_2
-\frac{1}{l}\right),
\label{JT_action}
\ee
where we have included the holographic boundary counter terms which makes the result on-shell finite. An  extensive treatment of holographic counter terms in 2d dilaton gravity can be found in \cite{Grumiller:2007ju}.  $I_{(0)}$ is given by
\be
I_{(0)}=-\frac{1}{8G_N}\int \sqrt{g_2}\,r_+\left(R_2-2\Lambda-\frac{Q^2}{2r_+^4}\right)-\frac{1}{4G_N}\int_\partial \sqrt{-\bar{\gamma}}\,r_+
\bar{K}
\label{topological_action_0}
\ee
where $Q=2r_+^2/\ell^2$ is fixed consistently by demanding that the metric $e.o.m.$  be satisfied. Thus we find
\be
I_{(0)}=-\frac{1}{8G_N}\int \sqrt{g_2}\,r_+ R_2-\frac{1}{4G_N}\int_\partial \sqrt{-\bar{\gamma}}\,r_+
\bar{K}
\label{topological_action}
\ee
as expected \cite{Jensen:2016pah,Maldacena:2016upp}. The JT action \eqref{JT_action} has been shown to describe the near horizon fluctuations of the BTZ metric near extremality for the charge $Q$ held constant. The $eom$ for the JT action constrains the metric to be locally $AdS_2$ while the dilaton $\phi$ solves a differential equation with the matter stress-tensor -if present, as the source.
\bea
&& R_2+\Lambda_2=0\cr&&\cr
&&\nabla_\mu\nabla_\nu\phi-g_{\mu\nu}\nabla^2\phi+g_{\mu\nu}\phi=4\pi G_N T^{(M)}_{\mu\nu}=0
\label{JT_eom}
\eea
\section{Near horizon near extremal limits of the BTZ}
We next would like to analyse the possible near horizon limits for near extremal BTZ. We first analyse such limits common in literature and revise the reasoning behind the JT model about such a near horizon geometry being able to capture the near extremal thermodynamics of BTZ when its angular momentum is held fixed. We then provide a one parameter family (parametrized by $\Ll$) of such limits and argue how it captures the near extremal thermodynamics of near extremal BTZ with a different combination of charges  held constant.
\subsection{Near extremal near horizon limit of  the BTZ}
The first law of black hole thermodynamics for the BTZ reads
\be
\delta M-\mu\,\delta J=T_H\,\delta\mathcal{S},
\label{first_law}
\ee
with $\delta J=0$ this then corresponds to
\be
\delta J=0\implies\delta M=T_H\delta\mathcal{S} 
\label{first_law_fixed_J}
\ee
Here the mass, angular momentum, entropy and the temperature of the BTZ are given by
\be
8G_N M=\rpl^2+\rmi^2,\hspace{0.2cm}8G_N J=2\rpl\rmi,\hspace{0.2cm}4G_N\mathcal{S}=2\pi \rpl,\hspace{0.2cm}2\pi T_H=\frac{\rpl^2-\rmi^2}{\rpl}
\label{BTZ_M_J_S_T}
\ee
The JT action \eqref{JT_action} captures the first law for $\delta J=0$ as follows: The 2d metric and the dilaton are constrained by the JT $e.o.m.$ in the absence of matter fields.
The 2d metric describes departures from the extremal limit and therefore has a horizon corresponding to a temperature $T^{(2)}_H$. The form of this 2d metric is that of an $AdS_2$ black hole with a temperature and can be obtained from taking the simultaneous near horizon near extremal limit\footnote{There does exist a different near horizon near extremal limit for the BTZ and we refer the readers to \cite{deBoer:2010ac} for the analysis.} of the BTZ metric \eqref{BTZ_metric_wiki} as in \cite{Gupta:2008ki}.
\be
\rpl-\rmi=2\lambda\,\delta \rpl  ,\,\,\,r= \rpl+\lambda(R-\delta \rpl),\,\,t=\frac{ T}{4 \lambda},\,\,\,\phi=y+\frac{1}{4\lambda}\left(1-\frac{2\lambda\,\delta\rpl}{\rpl}\right)T
\label{Sens_limit_original}
\ee
yielding the well known self-dual orbifold\footnote{This is an interesting solution to the Einstien's equation with negative cosmological constant with a null boundary and thus the boundary theory can be thought of as a discrete light cone quantized (DLCQ) field theory \cite{Balasubramanian:2009bg}. } 
\be
ds^2_{NH}=\frac{1}{4}\left[\frac{dR^2}{(R^2-\delta \rpl^2)}-(R^2-\delta \rpl^2) dT^2\right]+\rpl^2\left(dy^2+\frac{(R-\delta \rpl)}{2\rpl} dT\right)^2
\label{Sens_NH_metric_original}
\ee
here the 2d metric obeying the JT $eom$ is  given by the terms in the box bracket above in $\{R,T\}$ coordinates. Here the the 2d black hole has a temperature $T_{H}^{(2)}$ which has been set to $\delta\rpl/(2\pi)$. This 2d temperature must be related to the infinitesimal temperature $T_H$ of the BTZ. This is done by demanding that the periodicity close to the horizon of the Eclideanized time coordinate for the 2d metric and that of the BTZ geometry must be the same. To this end we note that the above scaling limit has a relative $1/4$ factor relating the BTZ time coordinate $t$ and the near horizon time $T$. We can absorb this factor by rescaling $R\rightarrow R/4$ and $\delta\rpl\rightarrow \delta\rpl/4$ $i.e.$ we work instead with the scaling
\be
\rpl-\rmi=\lambda\,\delta \rpl/2  ,\,\,\,r= \rpl+\lambda(R-\delta \rpl)/4,\,\,t=\frac{ T}{ \lambda},\,\,\,\phi=y+\frac{1}{\lambda}\left(1-\frac{\lambda\,\delta\rpl}{2\rpl}\right)T
\label{Sens_limit_modified}
\ee
which yields the same near horizon metric as above. Therefore one can now relate the temperatures of the 2d near horizon metric and that of the BTZ $i.e.$ $T^{(2)}_H=T_H$.   
One then solves for the $eom$ constraining the dialton \eqref{JT_eom} in such a 2d geometry. The solutions to the \eqref{JT_action} take the form
\bea
ds^2&=&\frac{dR^2}{(R^2-(2\pi T_H)^2)}-(R^2-(2\pi T_H)^2)dT^2\cr&&\cr
\phi &=& c_1 R+\sqrt{R^2-(2\pi T_H)^2}\left( c_2 e^{(2\pi T_H) T} + c_3e^{-(2\pi T_H) T }\right)
\eea
As one is interested in stationary solutions in verifying the first law one chooses the time invariant solution for $\phi$
\be
\phi= R\,\,\phi_\partial
\label{dilaton_0}
\ee
where $c_1=\phi_\partial$ determines the boundary value of the dialton. For the static case the solutions to the JT $eom$ are parametrized by the 2 constants: $T^{(2)}_H$ and $\phi_\partial$, here we have considered imposing Dirichlet boundary conditions on the metric and the dilaton. The temperature $T_H^{(2)}$ has been geometrically matched with that of the parent BTZ geometry. 
The Bekenstein-Hawking entropy of BTZ is always
\be
\mathcal{S}=\frac{\pi\rpl}{2G_N}.
\label{entropy_BTZ}
\ee 
The value of the dilaton $\Phi$ at extremality is the extremal horizon radius $r_0$, while the value of the dilaton solving \eqref{JT_eom} in the background \eqref{Sens_NH_metric_original} at its ($AdS_2$) horizon gives the fluctuation of the horizon radius and thus the BTZ entropy above extremality.
\bea
&&\mathcal{S}_{ext}=\frac{2\pi r_0}{4 G_N},\hspace{0.3cm}\mathcal{S}^{(2)}=\frac{2\pi \delta r_+ \phi_\partial}{4 G_N}=\frac{(2\pi)^2 T_H^{(2)}\phi_\partial}{4G_N}=\frac{\pi^2 T_H}{4G_N}=\mathcal{S}-\mathcal{S}_{ext}\cr&&\cr
&&\hspace{1cm}\implies\phi_\partial=1/4
\label{JT_entropy}
\eea
where we have used $T^{(2)}_H=T_H$. Here $\mathcal{S}_{ext}$ is simply obtained by computing  the on-shell Euclidean action  of the topological term $I_{(0)}$ in \eqref{JT_action}.
\be
I_{(0)}^{(OS)}=-\mathcal{S}_{ext}
\ee
As the boundary value of dilaton $\phi_\partial$ and $T_H^{(2)}$ are both fixed the ADM mass, $M^{(2)}$ as seen by the JT theory can then be deduced by 1) either computing the on-shell Euclidean action or 2) by computing the renormalized stress-tensor. 
\\\\
1) The Euclidean on-shell JT action on a static solution of \eqref{JT_eom} with the above values of $T^{(2)}_H$ and $\phi_\partial$ gives
\be
\beta F^{(2)}=-\mathcal{S}_{ext}+\beta M^{(2)}-\mathcal{S}^{(2)}
\ee   
where $\mathcal{S}_{ext}$ is obtained from the on-shell value of the topological term in \eqref{JT_action} and matches the extremal entropy of the BTZ. The above method yields
\be
M^{(2)}=\frac{\pi^2 {(T^{(2)}_H)}^2\phi_\partial}{2G_N}=\frac{\pi^2 {(T_H)}^2}{8G_N}=M_{BTZ}-M_{ext}
\label{JT_ADM_Mass}
\ee
2) The renormalised Brown-York stress-tensor is given by differentiating the renormalized on-shell JT action on static solutions of \eqref{JT_eom} $w.r.t.$ the boundary metric of the $AdS_2$ metric ( given in the square box bracket in \eqref{Sens_NH_metric_original} ). This gives the same ADM mass as above. 
\\\\
The first law of  black hole mechanics for near extremal BTZ for processes which do not change the angular momentum of the black hole are seen to be reproduced by the ADM mass and entropy seen by the JT theory only after $T^{(2)}_H$ and $\phi_\partial$ have been fixed as above. We see the first law for infinitesimal changes in $M^{(2)}$ and $\mathcal{S}^{(2)}$ given in \eqref{JT_ADM_Mass} \eqref{JT_entropy} by relating their variations
\be
\delta M^{(2)}=T_H\,\,\delta \mathcal{S}^{(2)}
\ee
where the changes in the JT mass and entropy in the above equation are the changes of the BTZ mass and entropy above extremality. 
\\\\
The excess mass above extremality is also rightly produced by the JT model. This can be easily seen by taking the ADM mass of the BTZ in terms of inner and outer horizons \eqref{BTZ_M_J_S_T} and scaling the horizons as \eqref{Sens_NH_metric_original}. However also note that the $\rpl$ in the limits \eqref{Sens_NH_metric_original} is the outer horizon of the near extremal solution $i.e.$ $\rpl=r_0+\lambda \delta\rpl$ where $r_0$ is the extremal horizon. We therefore find
\be
8G_N M=2\,r_0^2+2\lambda^2 \,\delta\rpl^2+\mathcal{O}(\lambda^2)\implies M-M_{ext}=\frac{\pi^2 T_H^2}{8G_N}=M^{(2)}
\ee
The same is true for the entropy of near extremal BTZ as the value of $\phi_\partial$ was chosen precisely for this \eqref{JT_entropy}. The fact that the angular momentum remains constant is implied by the above limits. 
\\\\
The value of the dilaton $\phi$ as seen in the JT model is completely determined by its boundary value and the its $eom$ \eqref{JT_eom} after the temperature of the 2d metric is matched appropriately with that of the parent near extremal BTZ. This dilaton $\phi$ given by \eqref{dilaton_0} is the linear fluctuation of the full dilaton $\Phi=r$  in the BTZ metric \eqref{BTZ_metric_wiki} over the extremal horizon value of $r=r_0$. The value of the $\Phi$ at the near extremal horizon as deduced from \eqref{BTZ_metric_wiki} and the scaling of the parameters as in \eqref{Sens_limit_modified} is 
\be
\Phi=\rpl=r_0+\delta\rpl/4
\ee 
which matches the value of the dilaton $\phi$ in the JT model at the horizon of the 2d metric
\be
\phi=\delta \rpl \phi_\partial\hspace{0.4cm}{\rm with} \,\,\,\phi_\partial=1/4
\ee
This further corroborates the match between the near horizon fields in the JT model and the higher dimensional geometry. Note had we not absorbed the $1/4$ in the scaling \eqref{Sens_limit_original} into the redefinition of $R$ and $\delta\rpl$ and chosen $T^{(2)}_H=T_H$ we would have run into inconsistency between the value of $\phi_\partial$ and the correct dependence of $\S-\S_{ext}$ on $T_H$.  The readers are referred to \cite{Johnstone:2013ioa} for a study of low temperature expansion of the first law of black hole thermodynamics for near extremal geometries.  
\subsection{New near horizon near extremal limits}
We next construct new near horizon limits by taking a cue from the form of the BTZ metric used to solve the Dray-'tHooft problem for rotating shockwaves \cite{Malvimat:2021itk}. Consider a dust of light particles released from the  boundary of the $AdS$ which slowly begin to fall towards the outer horizon, we consider this dust to have an angular momentum per unit energy  $\Ll$ and its trajectory would be traced by rotating in-falling geodesics. We can consider these to be massive but at very late times as they approach the black hole outer horizon their trajectory can be approximated by null geodesics as they would be constituting a shockwave. The analysis of the null geodesics limits the value of $\Ll$ to be $\leq 1$ \cite{Cruz:1994ir}. Here we only consider cases with $\Ll<1$\footnote{It is known that null geodesics with an $\Ll=\mu^{-1}$ for higher dimensional black holes would not reach the outer horizon (or even the ergosphere), here $\mu_{ext}=1$. Therefore we consider the $\Ll=1$ case to be special only for the BTZ. Further for $\Ll=1$ the mass and angular momentum deformations maintain extremality.}.
It is for such rotating shockwaves that the finely tuned  mutual information between the 2 CFTs constituting the TFD state dual to a rotating BTZ showed a late time scrambling at a rate given by a $\lambda_L>2\pi/\beta$ \cite{Malvimat:2021itk}. We also find that the blueshift suffered by the rotating null shockwave is governed by $\frac{2\pi}{\beta(1-\mu\,\Ll)}\geq\frac{2\pi}{\beta}$. Such a dust of in falling matter would perceive the BTZ metric along proper (affine) coordinates along its trajectory. As we would be interested in the near horizon region we would work with affine parametrization of the relevant rotating geodesic at the outer horizon.      
\\\\ 
We first begin with recasting the BTZ metric using the null geodesics with arbitrary angular momentum. 
We write the metric  line element squared \eqref{BTZ_metric_wiki} along in-out going null geodesics $\xi_\pm^\mu\partial_\mu$ as
\be
ds^2=F(\xi_+\cdot dx)(\xi_-\cdot dx)+h(dz+h_\tau d\tau)^2
\ee
where $\tau=t-\mathcal{L}\phi$, we define $z$ below later in \eqref{Kruskal_transverse_coord}. We define the vector $\xi^\mu\partial_\mu$ along  null geodesic with energy $\mathcal{E}=1$ and angular momentum $\mathcal{L}$ defined along the killing vectors of the rotating geometry $\zeta_E=\partial_t$ and $\zeta_L=\partial_\phi$ respectively $i.e.$
\bea
&&\xi^2=0,\,\,\,g_{\mu\nu}\xi^\mu \zeta_E^\nu=\mathcal{E}=1,\,\,\,g_{\mu\nu}\xi^\mu \zeta_L^\nu=\mathcal{L}\cr&&\cr
\implies&&\xi_\pm\cdot dx=\frac{r \sqrt{\mathcal{L} \left(\rmi^2 \mathcal{L}-2 \rmi \rpl+\rpl^2 \mathcal{L}\right)-r^2 \left(\mathcal{L}^2-1\right)}}{\left(r^2-\rmi^2\right) \left(r^2-\rpl^2\right)}dr  \pm (dt-\mathcal{L}d\phi)
\label{null_geodesic_0}
\eea  
This form of the metric is not unusual.
For instance for $\mathcal{L}=0$ we have
\be
ds^2=F \left[\frac{r^4 dr^2}{\left(r^2-\rmi^2\right)^2 \left(r^2-\rpl^2\right)^2}-dt^2\right]+\frac{\left(r^2 d\phi - \rmi \rpl dt \right)^2}{r^2}
\ee
where $F=\frac{\left(r^2-\rmi^2\right) \left(r^2-\rpl^2\right)}{r^2 }$.
For arbitrary $\mathcal{L}$ one can think in terms of Kruskal coordinates along the in-out going null geodesics. We begin with defining light-cone coordinates 
\be
du=\xi_-\cdot dx,\,\,\,\,\,dv=\xi_+\cdot dx.
\ee
One can integrate the above relation to find
\be
u=r_*-\tau,\,\,\,\,\,v=r_*+\tau,
\hspace{0.3cm}{\rm where}\,\,\,r_*(r')=-\int_\infty^{r'}\frac{r \sqrt{\mathcal{L} \left(\rmi^2 \mathcal{L}-2 \rmi \rpl+\rpl^2 \mathcal{L}\right)-r^2 \left(\mathcal{L}^2-1\right)}}{\left(r^2-\rmi^2\right) \left(r^2-\rpl^2\right)}dr  
\label{light_cone_coords}
\ee
It can be seen that the vector fields along the above light-cone coordinates are not affine at the outer horizon, we can therefore 
define  affine coordinates $\{U,V\}$ at the horizon  for generic $\mathcal{L}$ as follows
\bea
&&U=-e^{\kappa \,u},\,\,\,\,V=e^{\kappa\, v},\,\,{\rm where}\,\,\,\kappa
=\frac{\rpl(1-\mu^2)}{(1-\mu\,\mathcal{L})}=\frac{2\pi}{\beta(1-\mu\,\mathcal{L})}
\label{affine_horizon}
\eea
The vector field along these coordinates $\chi_-=\partial_U\,\, \chi_+=\partial_V $  satisfies $\chi_\pm^\mu\nabla_\mu\chi_\pm^\alpha=0$ (affine condition).  Therefore the value of $\kappa$ is a result of demanding smoothness along such rotating Kruskal coordinates.
The metric in these coordinates takes the form
\be
ds^2=\frac{F}{\kappa^2 UV}dUdV +h\left(dz+h_\tau\frac{UdV-VdU}{2\kappa UV}\right)^2
\label{Kruskal_L}
\ee
The coordinates $\{U,V\}$ see the  future and past horizons at $V=0$ and $U=0$ in the right exterior respectively. The left exterior is obtained by reversing their signs. The $z$ coordinate is defined  by demanding that $\frac{h_\tau}{UV}$ is finite on either of the horizons. This implies
\bea
&&z=\phi-\mu t,\,\,\,\,\,\tau=t-\mathcal{L}\phi,\,\,\,{\rm where}\,\,\,\mu=\frac{\rmi}{\rpl}\cr&&\cr
&\implies&\phi=\frac{z+\mu \tau}{1-\mu\mathcal{L}},\,\,\,\,\,t=\frac{\tau+\mathcal{L}z}{1-\mu\mathcal{L}}
\label{Kruskal_transverse_coord}
\eea
The scaling of $z$ $w.r.t$ $\phi$ can be fixed by demanding that the horizon area remains the same in $\{U,V,z\}$ coordinate for $z\in \{0,2\pi\} $. This also in turn implies that the $z$ coordinate is the co-rotating coordinate at the horizon.
The conical deficit as seen by the (Euclidean) $g_{UV}$ part of the metric \eqref{Kruskal_L} is $\kappa$. Note that $\kappa$ corresponds to the temperature of the BTZ only for $\mathcal{L}=0$, else it is greater than later.
\subsubsection{Near horizon family for $\mathcal{L}<1$} In what follows we would consider the above metric \eqref{Kruskal_L} in Kruskal coordinates parametrized by $\Ll$ but in $\{r,\tau,z\}$ coordinates.
\be
ds^2=F \left(\frac{dr^2}{f^2}-d\tau^2\right) +h\left(dz+h_\tau d\tau\right)^2
\label{Kruskal_L_1}
\ee
where $dr_*=dr/f$.
Note, that $\Ll$ parametrizes the relation between the non-radial coordinates $\{\tau,z\}$ and $\{t,\phi\}$. 
\\\\
For $\mathcal{L}=0$ we get $z=y$ $i.e.$ the $z$ coordinate required to have a smooth metric at the horizon without any coordinate singularity. Following the simultaneous scaling limit \eqref{Sens_limit_original}  we get \eqref{Sens_NH_metric_original}
\be
ds^2_{NH}=ds^2_{NH_{\cL= 0}}=\frac{1}{4}\left[\frac{dR^2}{(R^2-\delta\rpl^2)}-(R^2-\delta\rpl^2)dT^2\right]+\rpl^2\left(dy^2+\frac{(R-\delta\rpl)}{2\rpl}dT\right)^2
\label{Sens_NH_metric}
\ee
with the recognizable $AdS_2$ factor with a $T^{(2)}_H$ determined by $\delta\rpl$. For generic value of $\mathcal{L}<1$ we work with the metric \eqref{Kruskal_L_1} in the coordinates $\{r,\tau,z\}$ and scale the coordinates and parameters as
\be
\rpl-\rmi=2\lambda\delta\rpl(1-\Ll)/4,\,\,\,r= \rpl+\lambda \tfrac{(1-\Ll)}{4} (R-\delta\rpl),\,\,\tau=\frac{T}{ \lambda}
\label{Sens_limit_L}
\ee
Note that the $z$ coordinate needs no further scaling as it always ensures the smoothness of the line element transverse to the 2d metric line element.
We therefore find a family of identical near horizon metrics of the form \eqref{Sens_NH_metric} but with $y\rightarrow z$.
\be
ds^2_{NH_\Ll}=\frac{1}{4}\left[\frac{dR^2}{(R^2-\delta\rpl^2)}-(R^2-\delta\rpl^2)dT^2\right]+\rpl^2\left(dz^2+\frac{(R-\delta\rpl)}{2\rpl}dT\right)^2
\label{NH_metric_L}
\ee 
 Therefore for any value of $\mathcal{L}<1$ we have the same near horizon metric as given by \eqref{Sens_NH_metric} with $\{\tau,z\}$ given by \eqref{Kruskal_transverse_coord}. 
It is important to recognize that the above metric differs from \eqref{Sens_NH_metric} as the coordinate transverse to the $AdS_2$ factor $z$ is related differently to the 3d BTZ metric coordinates as compared to $y$  \eqref{Kruskal_transverse_coord}. 
Further the time coordinates $\tau$ and $t$ -although scaled identically for obtaining the near horizon metric, are also related by \eqref{Kruskal_transverse_coord} parametrized by $\Ll$. 
Like the $\mathcal{L}=0$ case, we expect  the JT model about the above metric for arbitrary values of $0\leq\mathcal{L}<1$ would capture the thermodynamics of the parent near extremal BTZ for a unique relation between $T^{(2)}_H$ and the infinitesimal near extremal BTZ temperature.
\\\\
The above metric can be understood as obtaining a near horizon near extremal metric by approaching the outer horizon along null geodesics with unit energy and finite angular momentum $0\leq\Ll<1$. This can be seen by relating the  Kruskal coordinates of the AdS$_2$ factor of the above metric 
\be
dU^{(2)}=\frac{dR}{(R^2-\delta\rpl^2)}-dT,\hspace{0.2cm}dV^{(2)}=\frac{dR}{(R^2-\delta\rpl^2)}+dT
\ee 
to $\{U,V\} $ in the $\lambda\rightarrow 0$ limit. Using the limit \eqref{Sens_limit_L} on the  coordinates \eqref{light_cone_coords} we find
\be
U\rightarrow (U^{(2)})^{1/\lambda},\hspace{0.3cm}V\rightarrow (V^{(2)})^{1/\lambda}
\label{Kruskal_to NH_Kruskal}
\ee
thus the regions close to the outer horizon in $\{U,V\}$  coordinates ($U=0=V$) get mapped to the entire domain of $\{U^{(2)},V^{(2)}\}$ coordinates as $\lambda\rightarrow 0$\footnote{The relation between $\{U,V\}$ and$\{U^{(2)},V^{(2)}\}$ is for their absolute values. The $-$ve signs for the exterior regions are restored after taking the limit.}. Thus taking the $\lambda\rightarrow 0$ limit amounts to going near horizon along in-out null geodesics parametrized by the $\{U,V\}$ coordinates. The fact that the metric \eqref{Kruskal_L} is smooth along such a limit is ensured by the coordinate transformations \eqref{Kruskal_transverse_coord} and \eqref{affine_horizon}. Equivalently it can also be easily seen that the Euclidean form of the metric \eqref{Kruskal_L_1} would be smooth at the outer horizon if its time coordinate has a periodicity of $2\pi/\kappa$ \eqref{affine_horizon}.
\\\\
The above form of the near horizon metric \eqref{NH_metric_L} can be thought of as being encountered by null in-falling particles with an angular momentum per unit energy $\Ll$ as they approach the near horizon region. One could also consider massive particles released from the BTZ boundary with similar angular momentum as its time-like geodesics would tend to null geodesics as the particles get blue shifted while approaching the outer horizon.  
\\\\
It was apparent while arriving at the JT model form the Einstein-Hilbert action in $AdS_3$ about the near extremal BTZ that the charge $Q$ \eqref{F_eom} associated with the gauge field due to dimensional reduction is held fixed. This electric charge seen in the 2d picture is the space-time charge associated with the translations in the direction which got dimensionally reduced $i.e.$ the $y$-direction in \eqref{dim_red_metric}. It is easy to see that for the usual $\Ll=0$ case this corresponds to $Q=J$. For the case with $\Ll<1$ this charge $Q$ would correspond to the BTZ charge associated with translations in $z$ which is
\bea
Q=\mathcal{Q}[\zeta]=\frac{J-\Ll\, M}{\sqrt{1-\Ll^2}},\hspace{0.4cm}
{\rm where}\,\,\zeta=\frac{1}{\sqrt{1-\Ll^2}}(\partial_\phi+\Ll\partial_t)
\eea
here $\zeta$ is the normalized vector field proportional to $\partial_z$ which generates translations in the $z$ direction. Therefore the JT model obtained about the near horizon metric \eqref{NH_metric_L} describes the near extremal thermodynamics with 
\be 
\delta Q\sim\delta J-\Ll\delta M=0\implies\delta J=\Ll\,\delta M
\label{charge_constraint}
\ee
Plugging the above constraint in the first law of black hole thermodynamics \eqref{first_law} we expect  the relation
\be
\delta M=\frac{T_H}{(1-\mu\,\Ll)}\delta \mathcal{S}
\label{first_law_L_finite}
\ee
to be captured by the JT model in this case. We first note in the next section how does the mass $M$ and entropy $\mathcal{S}$ of the BTZ change with respect to its temperature when an extremal configuration is perturbed according to the above constraint in \eqref{charge_constraint}. 
\section{Thermodynamics}
In this section we would like to describe the  thermodynamics of the configurations captured by the near extremal limit obtained in the previous section for $0\leq\Ll<1$ and then compare it systematically with the relevant thermodynamics of near extremal configurations of near extremal BTZ. The analysis in this section would mimic the consistency checks described in the previous section for the case of $\Ll=0$ $i.e.$ the JT model describing near extremal BTZ found till now in literature. To differentiate with the JT description for the $\Ll=0$ case we call the JT model for $0\leq \Ll<1$ as the JT$_{\Ll}$ model. Thus the JT$_{\Ll=0}=$ JT model for near extremal BTZ.
\\\\
We first calculate the mass and entropy over extremality for near extremal BTZ when an extremal configuration is changed with the addition of $M_{\rm ext}\rightarrow M_{\rm ext}+\Delta M$ and $J_{\rm ext}\rightarrow J_{\rm ext}+\Delta J$ such that $\Delta J=\Ll\,\Delta M$. We then compute the excess mass and entropy in the equivalent JT$_\Ll$ description and show a match. The reason we consider such a relation between $\Delta J$ and $\Delta M$ is that the near horizon form of the metric   \eqref{NH_metric_L} is obtained from re-writing the BTZ metric along in-out going null geodesics with angular momentum per unit energy $\Ll$ \eqref{Kruskal_L} or \eqref{Kruskal_L_1}. Thus any perturbation to the geometry would imply the above relation between $\Delta J$ and $\Delta M$. 
\subsection{BTZ} 
We consider extremal BTZ and then consider adding a mass $\Delta M=M-M_{ext}$ and angular momentum $\Delta J=J-J_{ext}$ to it such that $\Delta J=\Ll\Delta M$. The inner and outer horizons of the BTZ geometry in terms of its charges and its near extremal limit are
\bea
&&r_\pm=\frac{\sqrt{M+J}\pm\sqrt{M-J}}{2}\cr&&\cr
\implies&& r_\pm=r_0+\delta r_\pm=\frac{\sqrt{2M}}{2}\pm\frac{1}{2}\sqrt{\Delta M(1-\Ll)}+\mathcal{O}(\Delta M)
\label{BTZ_delta_rpm}
\eea
Similarly noting the infinitesimal temperature gained to be
\bea
&&2\pi T_H=2(\delta \rpl-\delta\rmi)=2\sqrt{8G_N}\sqrt{\Delta M(1-\Ll)}
\eea
Therefore we find the excess mass and the entropy of the BTZ to be
\be
8G_N\Delta M=\frac{\pi^2 T^2_H}{(1-\Ll)},\hspace{0.5cm}\Delta\mathcal{S}=\mathcal{S}-\mathcal{S}_{ext}=\frac{\pi^2 T_H}{4G_N}
\label{Delta_mass_entropy_temp}
\ee
The first law \eqref{first_law_L_finite} infinitesimally  close to extremality is then recovered by expressing the above excess mass $\Delta M$ and entropy $\Delta S$ in terms of infinitesimal temperature rise $\delta T_H$ as
\be
\delta M=\frac{T_H}{(1-\Ll)}\delta \mathcal{S}
\label{first_law_L_ext}
\ee
For $\Ll=0$ we seamlessly recover the case found in literature. The above form of the first law must be considered as a near extremal expansion of $\eqref{first_law_L_finite}$ as
\be
\frac{T_H}{(1-\mu\,\Ll)}=\frac{T_H}{(1-\Ll)}\left[1+\frac{(1-\mu)\Ll}{(1-\Ll)}+\mathcal{O}((1-\mu)^2)\right]
\ee
It is the above given excess mass and entropy \eqref{Delta_mass_entropy_temp} and the first law \eqref{first_law_L_ext} close to extremality that we hope to recover from the JT description in the near horizon region about the metric \eqref{NH_metric_L}.
\subsection{JT$_\Ll$ description}
The equivalent JT description is obtained by analysing the JT$_\Ll\,$ model about the near horizon metric \eqref{NH_metric_L} obtained from the BTZ metric written about the in-out going rotating null coordinates \eqref{Kruskal_L} parametrized by rotation parameter $\Ll$.
Therefore the action for the JT$_\cL$ model is the same as \eqref{JT_action} obtained about a near horizon metric \eqref{NH_metric_L}. Thus given the relations \eqref{Kruskal_transverse_coord} the coordinates of the JT$_\cL$ model are related to boundary coordinates differently (parametrized by $\cL$) as compared to those in the JT model. 
Just as in the usual ($\Ll=0$) case we would first like to match the temperature of the near horizon 2d metric obtained in \eqref{NH_metric_L} $i.e.$
\be
ds^2_{JT_\cL}=\frac{dR^2}{R^2-\delta\rpl^2}-(R^2-\delta\rpl^2)dT^2
\label{JT_metric_L}
\ee
with that of the parent BTZ metric. 
We first note the form of the BTZ metric which is used to obtain the near horizon geometry \eqref{NH_metric_L} in Kruskal coordinates
\be
ds^2=\frac{F}{\kappa^2 UV}dUdV +h\left(dz+h_\tau\frac{UdV-VdU}{2\kappa UV}\right)^2=F \left(\frac{dr^2}{f^2}-d\tau^2\right) +h\left(dz+h_\tau d\tau\right)^2
\label{Kruskal_L_2}
\ee
The periodicity of the Euclidean $T$ coordinate in \eqref{JT_metric_L} close to its horizon at $R=\delta\rpl$ should therefore match that of $\tau$ coordinate close to the outer horizon in the above form of the metric. Determining the later implies demanding that the Euclidean geometry in the $\{r,\tau\}$ directions above resemble a disk without a conical deficit. This in turn is equivalent to defining the smooth or affine coordinates $\{U,V\}$ at the out horizon. The periodicity of the $\tau$ coordinate therefore is $2\pi/\kappa$ where 
\be
\kappa=\frac{2\pi}{\beta (1-\mu\Ll)}
\ee
as determined by demanding smoothness of in-out going Kruskal coordinates $\{U,V\}$ \eqref{affine_horizon}. We therefore have the temperature of the 2d geometry defined as 
\be
2\pi T^{(2d)}_H=\delta\rpl=\kappa=\frac{2\pi T_H}{(1-\mu\,\Ll)}\overset{\mu\rightarrow 1}{\xrightarrow{\hspace{1cm}}}\frac{2\pi T_H}{1-\Ll}
\label{JT_temp_L}
\ee
Note that the above method is equivalent to the geometric reasoning used for the usual ($\Ll=0$) case seen thus far in literature. Equivalently, any other relation between  $T^{(2d)}_H$ and $T_H$ would create a mismatch and amount to  introducing a coordinate singularity at the out horizon. 
\\\\
We next turn to determining the value of the excess entropy. We first note the value of the dilaton in JT$_\Ll$ model 
\be
\phi = c_1 R+\sqrt{R^2-\delta\rpl^2}\left( c_2 e^{\delta\rpl T} + c_3e^{-\delta\rpl T }\right)
\label{JT_dilaton_full_L}
\ee
which for the static case implies the same form as \eqref{dilaton_0}
\be
\phi=R\,\,\phi_\partial
\label{JT_dilaton_L_static}
\ee
where $\phi_\partial$ being its regularised boundary value. The value of the dilaton in the JT model at the horizon captures the excess entropy over extremality, we therefore have
\bea
&&\mathcal{S}-\mathcal{S}_{ext}=\mathcal{S}^{(2)}=\frac{2\pi\delta\rpl\phi_\partial}{4G_N}=\frac{\pi^2 T^{(2d)}_H\phi_\partial}{G_N}=\frac{\pi^2 T_H}{4G_N}\cr&&\cr
\implies && \phi_\partial=(1-\Ll)/4
\label{Delta_entropy_L}
\eea
Like in the $\Ll=0$ case we can ascertain that the above value of dilaton in the JT$_\Ll$ model is consistent with the full value of the dilaton $\Phi$ obtained from the BTZ metric \eqref{Kruskal_L_1} where $\Phi=\sqrt{h}$. Expanding $\Phi=\sqrt{h}$ using the scaling \eqref{Sens_limit_L} and further expanding the near extremal $\rpl$ as\footnote{This value of $\delta\rpl$ is not the same as the one in \eqref{BTZ_delta_rpm}, we know however that the change in the inner and outer horizons is equal in  magnitude and opposite in sign even when $\Delta J=\Ll \Delta M$ close to extremality.}
\be
\rpl=r_0+\delta\rpl (1-\Ll) \lambda/4
\ee
we have
\be
\Phi|_{\rpl}=\sqrt{h}|_{\rpl}=r_0+\delta\rpl(1-\Ll)/4=r_0+\phi|_{\delta\rpl}
\ee
wherein we expand to linear order in $\lambda$ and put $\lambda=1$. The excess value over the extremal horizon $\Phi=r_0$ matches the static value of the dilaton in the JT$_\Ll$ model \eqref{JT_dilaton_L_static} at the 2d horizon $R=\delta\rpl$ for $\phi_\partial=(1-\Ll)/4$. 
Therefore for the above regularised boundary value of the dilaton the JT model passes the same check as in the $\Ll=0$ case and reproduces the right near extremal excess entropy as a function of the infinitesimal temperature.
\\\\
The topological part of the JT action \eqref{topological_action} yields (minus) the extremal entropy of BTZ for the  2d metric \eqref{JT_metric_L} exactly as in the $\Ll=0$ case. Here one has to use the periodicity of the Euclidean time coordinate as dictated by $\delta\rpl$. As the final answer is independent of the 2d temperature this value is the same as in $\Ll=0$ case.
\be
\mathcal{S}_{ext}=\frac{\pi\rpl}{2G_N}
\ee 
We next turn to determining the ADM mass in the JT$_\Ll$ description having fixed both the $T^{(2)}_H$ and $\phi_\partial$. Here again we can determine the mass $M^{(2)}$ in two ways: 1) either by the bulk on-shell action 
\be
\beta^{(2)}F^{(2)}=\mathcal{S}_{ext}+\beta^{(2)}M^{(2)}-\mathcal{S}^{(2)}
\ee
where we use $\beta^{(2)}=(T^{(2)}_H)^{-1}$; (2) or by the renormalized Brown-York stress-tensor. Either ways the answer is the same as before with respect to $T^{(2)}_H$ and $\phi_\partial$
\be
M^{(2)}=\frac{\pi^2(T^{(2)}_H)^2\phi_\partial}{2G_N}=\frac{\pi^2(T_H)^2}{8G_N(1-\Ll)}=M_{BTZ}-M_{ext}=:\Delta M
\label{Delta_mass_L}
\ee
where we used \eqref{JT_temp_L} and $\phi_\partial=(1-\Ll)/4$. From \eqref{Delta_entropy_L} and \eqref{Delta_mass_L} we see that the excess mass and extropy over extremality of the parent BTZ geometry is reproduced exactly by the JT description. The first law \eqref{first_law_L_ext} is also reproduced. 
\\\\
Thus the match implies that the JT$_\Ll$ model correctly captures the near extremal thermodynamics of the BTZ geometry for processes where  $\Delta J=\Ll\, \Delta M$. The important peculiarity of the JT$_\Ll$ description is that the 2d near horizon  metric intrinsic to the JT$_\Ll$ analysis sees a temperature which is greater than that of the temperature of the near extremal BTZ.
\section{2d scalar in JT$_{\Ll}$  }
We next turn to the interaction of the above JT$_\Ll$ model  with matter fields along the lines of \cite{Jensen:2016pah} and \cite{Maldacena:2016upp}. The JT theory was shown to rightly reproduce the near extremal thermodynamics of a large class of black holes in $AdS_{3,4,5}$ and  flat space-times but with all global space-time charges -except its mass, held constant \cite{Nayak:2018qej,Castro:2018ffi,Moitra:2018jqs,Moitra:2019bub}. This -as we have seen the previous section for the case of BTZ, invariably implied that the $AdS_2$ metric of the JT theory has a temperature which is identified with temperature $T_H$ of the parent near extremal BH. As a consequence the 4pt OTOC of 2d probe scalars in such a JT theory sees a Lyapunov index which is $2\pi T_H$  with the Schwarzian action governing the interaction between the scalars and the JT metric. Given the findings in the previous section it is worth revisiting this in the 2d setting of \cite{Jensen:2016pah,Maldacena:2016upp} for the case of the JT$_\Ll$ theory.
\\\\
Ideally we imagine the 2d scalar fields resulting from the dimensional reduction of 3d matter fields in the near extremal BTZ background. 
Any null in-falling matter with an angular momentum per unit energy $0\leq\Ll<1$ would end up seeing the $AdS_2$ geometry mentioned in section 3.
Equivalently one can dimensionally reduce along transverse coordinate $z$ using the metric \eqref{Kruskal_L_1} for dimensional reduction. Note that the relation between the non-radial coordinates used in the near horizon analysis and the Boyer-Lindquist coordinates  are related $via$ $\Ll$ \eqref{Kruskal_transverse_coord}.
However we do not show this reduction from 3d to 2d explicitly here. We only consider scalar fields in the 2d AdS JT$_\Ll$ theory for a fixed $\Ll$.  

The interaction of the JT$_\Ll$ theory with scalar matter fields is modelled to capture the essentials of the near extremal dynamics of AdS black hole geometries with the higher dimensional scalar fields. The JT theory we described above for $\Ll<1$ has its 2d metric and the dilaton configuration given by \eqref{JT_metric_L} and \eqref{JT_dilaton_full_L} respectively where the only difference from the $\Ll=0$ case is that the 2d black hole horizon relates to the infinitesimal BTZ temperature as		
\be
\delta\rpl=\frac{2\pi T_H}{1-\mu\,\Ll}=2\pi T_H^{(2)}.
\label{2d_horizon_BTZ_temp}
\ee
We saw in the previous section that such a match is necessary in order for the JT$_\Ll$ model to capture the near extremal thermodynamics of BTZ when its angular momentum and mass change as $\Delta J=\Ll\Delta M$. If -following the analysis of \cite{Jensen:2016pah,Maldacena:2016upp}, we consider a 2d scalar field which couples minimally to the 2d metric but not to the dilaton $i.e.$
\be
S_{matter}=\int d^2x \sqrt{-g} \,(\nabla \psi)^2
\ee
we immediately see that in the 2d metric given  by \eqref{JT_metric_L} the 2pt function for the primary operator $W$ in the dual CFT$_1$ is given by
\be
\langle W(\tau_1)W(\tau_2)\rangle=\frac{P}{\sinh^2\left(\pi T^{(2)}_H(\tau_1-\tau_2)\right)}
\ee
where $P$ is the 2pt. coefficient. 
One can then consider the case where 2 probe scalar fields $\psi$ and $\chi$ interact in the bulk only gravitationally. 
\be
S_{matter}=\int d^2x\sqrt{-g}\,\left((\nabla\psi)^2\,+\,(\nabla\chi)^2\right)
\ee 
and are dual to the CFT$_1$ operators $W$ and $V$ respectively\footnote{One needs to add boundary terms to the matter action for on-shell renormalizability.}. The procedure for obtaining the corrections to the 4pt function $\bra WW VV\ket$ in the small $G_N$ approximation in JT theory was done in \cite{Almheiri:2014cka,Jensen:2016pah,Maldacena:2016upp}. The 1d conformal transformations of the boundary CFT parametrized by $\tau\rightarrow f(\tau)$ correspond to bulk diffeomorphisms generating a family of bulk metric about \eqref{JT_metric_L} of the form
\bea
ds^2_{JT}&=&\frac{dr^2}{r^2}+\left(r^2+\frac{1}{2} F(\tau)
+\frac{1}{4r^2}F(\tau)^2\right)d\tau^2\cr&&\cr
F(\tau)&=&2\lbrace f,\tau\rbrace-\left(2\pi T_H^{(2)}f'\right)^2 
\label{JT_family_metric}
\eea
where we use $\{r,\tau\}$ coordinates and $\lbrace f,\tau\}$ is the Schwarzian derivative
\be
\lbrace f,\tau \rbrace=\frac{2f'f'''-3f''^2}{2f'^2}
\label{schwarzian}
\ee
The change in the 2pt functions of any boundary primary operator of weight $h$ is the given by
\be
\bra W(f(\tau_1))W(f(\tau_2))\ket=\left(\frac{\partial f(\tau_1)}{\partial\tau_1}\frac{\partial f(\tau_2)}{\partial\tau_2}\right)^{h/2}\bra W(\tau_1)W(\tau_2)\ket
\label{2pt_finite_change}
\ee
The on-shell bulk action for the JT theory \eqref{JT_action} evaluated on the metric \eqref{JT_family_metric} get contribution from the boundary term in the action as the bulk term vanishes. This on-shell action then provides the boundary effective action for the field $f$
\be
\S_{JT}^{(OS)}=-\frac{\phi_\partial}{8\pi G_N}\int d\tau \left[\lbrace f,\tau\rbrace +\tfrac{1}{2}\left(\pi T^{(2)}_Hf'\right)^2\right]
\label{JT_OS}
\ee
We next evaluate the change in the 2pt function \eqref{2pt_finite_change} for infinitesimal change in the $\tau$ coordinate $\tau= f(\tau)=\tau+\epsilon(\tau)$
\bea
\bra W(\tau_1)W(\tau_2)\ket &\rightarrow& \bra W(\tau_1)W(\tau_2)\ket \,\,\left[1 + \mathcal{B}(\epsilon(\tau_1),\epsilon(\tau_1)) \right]\cr&&\cr
 \mathcal{B}(\epsilon(\tau_1),\epsilon(\tau_2))&=&\Delta_W	\left[\epsilon'(\tau_1)+\epsilon'(\tau_2)-\frac{\epsilon(\tau_1)-\epsilon(\tau_2)}{\tan \left(\pi T^{(2)}_H (\tau_1-\tau_2)\right)}\right]
\label{B}
\eea
The $\epsilon$ operators are the boundary duals of the perturbative 2d graviton exchanges in the bulk. The correlators for the $\epsilon$s is then inferred from their quadratic action obtained by expanding \eqref{JT_OS} by substituting $f(\tau)=\tau+\epsilon(\tau)$ to quadratic order in $\epsilon$. 
\be
S_{\epsilon}=\frac{1-\Ll}{16\pi G_N}\int d\tau\left(\epsilon''^2-2\pi T^{(2)}_H\epsilon'^2\right)
\label{epsilon_action_quad}
\ee 
where we used the value of $\phi_\partial$ in the previous section \eqref{Delta_entropy_L}.
The $\epsilon$ operators can be recognised as the ``reparametrization modes'' of the $nAdS_2/nCFT_1$, the extensive formalism for which was developed in \cite{Haehl:2019eae} for arbitrary dimensions. 
The above action has zero modes for $\epsilon=\{1,e^{\pm 2\pi iT^{(2)}_H\tau}\}$ which correspond to the global $SL(2,R)$ symmetry in the bulk and global conformal symmetry in the CFT$_1$. The $\bra \epsilon (\tau_1)\epsilon (\tau_2)\ket$ is then computed on the thermal circle of periodicity $\beta^{(2)}=1/T^{(2)}_H$ excluding the contributions from the zero modes
\be
\bra \epsilon(\tau)\epsilon(0)\ket=\frac{4G_N}{1-\Ll}\left[-2\pi^2(T^{(2)}_H)^2\left(|\tau|-\tfrac{1}{2}\right)^2+4\pi^2 (T^{(2)}_H)^2\left(|\tau|-\tfrac{1}{2}\right)^2\sin 2\pi T^{(2)}_H|\tau| +c_1+c_2\cos2\pi T^{(2)}_H|\tau|\right].
\label{epsilon_propagator}
\ee
The order $G_N$ correction to the CFT$_1$ 4pt function is then given by 
\be
\frac{\bra W(\tau_1)W(\tau_2)V(\tau_3)V(\tau_4)\ket}{\bra W(\tau_1)W(\tau_2)\ket\bra V(\tau_3)V(\tau_4)\ket}=1+\bra \mathcal{B}\left(\epsilon(\tau_1),\epsilon(\tau_2)\right)\,\mathcal{B}\left(\epsilon(\tau_3),\epsilon(\tau_4)\right)\ket
\label{correction_0}
\ee
The above correlator is Euclidean and the time ordering on the thermal circle is different for time ordered versus out of time ordered due to the mod sign in \eqref{epsilon_propagator}. One can continue onto the Lorentzian time for either of the cases. The \emph{out of time ordered correlator} for large time separations $\frac{1}{2\pi T^{(2)}_H}\log\frac{ T_H}{G_N}\gg\delta t \gg \frac{1}{2\pi T^{(2)}_H}$ we have
\be
\frac{\bra W(0)V(\delta T)W(0)V(\delta t)\ket}{\bra W(0)W(0)\ket\bra V(\delta t)V(\delta t)\ket}\sim 1+\frac{8\pi G_N\Delta_W\Delta_V}{T_H}e^{2\pi T^{(2)}_H \delta t}
\label{OTO}
\ee
Therefore we see  the expected behaviour of the 2d scalar fields exhibiting a Lyapunov index given by $2\pi$ times the temperature of the 2d geometry 
\be
\lambda_L=2\pi T^{(2)}_H=\frac{2\pi T_H}{(1-\mu\,\Ll)}.
\ee
Thus we see that the rate of scrambling as seen by probe fields in such a JT theory is controlled by a Lyapunov index which depends on $\kappa$ given in eq\eqref{affine_horizon} obtained by demanding that the in falling null rotating geodesics be smooth at the outer horizon. This should not be surprising as we have already established in the previous section that the correct 2d temperature for the near horizon AdS$_2$ metric in the JT theory -such that it captures the relevant near extremal thermodynamics of near extremal BTZ with temperature $T_H$, is $T_{H}/(1-\mu\,\Ll)$.
\\\\
Here we considered scalar fields in the JT theory which do not couple to the dilaton \cite{Jensen:2016pah,Maldacena:2016upp}. The matter fields in the JT theory are also to be considered as the dimensionally reduced avatars of similar fields in the BTZ. Since the BTZ theory is more tractable than its higher dimensional counterparts one can indeed solve for $G_N$ or $1/c$ corrections to the 4pt functions of pair of primary operators in CFT$_2$ $\bra WWVV\ket$ dual to minimally interacting bulk fields in the BTZ. The gravitational interactions between the two probe scalars is dual to the stress tensor block in the CFT$_2$, this in the case of rotating BTZ implies that the left moving mode sees a left moving temperature $\frac{2\pi T_H}{(1-\mu)}$ while the  right moving mode sees a right moving temperature $\frac{2\pi T_H}{(1+\mu)}$. This corresponds to left moving and right moving Lyapunov indices \cite{Poojary:2018esz,Jahnke:2019gxr}. It was noted that imposing periodicity on the boundary spatial coordinate induces a sawtooth like behaviour in the OTOC of such a 4pt correlators \cite{Mezei:2019dfv}\footnote{In this case the periodicity was imposed on the co-moving $z$ coordinate.}. 
It would be interesting to understand how such a near horizon perspective squares with the analysis of \cite{Mezei:2019dfv,Craps:2020ahu,Craps:2021bmz} by analysing the Kaluza-Klein modes of the dimensionally reduced 3d probe scalar fields and their interactions with the fields in the JT$_\Ll$ theory.  We expect in this case the parameter $\Ll$ to be integrated out  in the 3d picture by coupling it to the appropriate Kaluza-Klein  modes of the 3d scalar fields. We leave this exercise for the near future.
\section{Conclusions \& Discussions}
In this paper we see that the JT theory describing the near horizon dynamics of near extremal BTZ is more capable at describing the departures form extremality than previously explored in literature $i.e.$ the constant angular momentum departures from extremality. This is made possible by writing the metric in terms of in-out going null geodesics with angular momentum per unit energy $\Ll$ \eqref{Kruskal_L} first used in \cite{Malvimat:2021itk} and then taking a simultaneous near extremal near horizon limit generalizing similar previously known limits \cite{Gupta:2008ki,Sen:2008yk}. The thermodynamics of the JT theory obtained -which we refer to as JT$_\Ll$; about such a near horizon metric parametrized by $0\leq\Ll<1$ is consistent with the near extremal thermodynamics of BTZ where the excess mass and angular momentum are related by $\Delta J=\Ll\,\Delta M$. We also verify that the behaviour of excess mass and entropy above extremality in the JT$_\Ll$ theory matches that of the relevant near extremal BTZ configuration. Thus, we also find that the expected form of the first law of black hole thermodynamics is reproduced by the near horizon description. A crucial requirement for this match in the thermodynamics is to identify the near horizon $AdS_2$ throat temperature as $T_H/(1-\mu\,\Ll)$. As a consequence we see that a computation of the 4pt OTOC of 2d probe scalar fields along the lines of \cite{Jensen:2016pah,Maldacena:2016upp} exhibits an exponential scrambling behaviour with a Lyapunov index $\lambda_L=2\pi T_H/(1-\mu\,\Ll)$. We therefore are able to see partially the near horizon or the IR dynamics which gives rise to the sawtooth  like behaviour in the OTOC in a state dual to rotating BTZ \cite{Mezei:2019dfv,Craps:2020ahu} and the scrambling of mutual information controlled by $\lambda_L>2\pi T_H$ \cite{Malvimat:2021itk}. This near horizon description also demonstrates how the generalized bound on the growth of chaos as derived in \cite{Halder:2019ric} can be viewed holographically. 
\\\\
The near horizon description used to obtain the JT theory in \cite{Jensen:2016pah} and \cite{Maldacena:2016upp} from higher dimensional gravity theories \cite{Nayak:2018qej,Moitra:2018jqs,Moitra:2019bub,Castro:2018ffi,Banerjee:2019vff} analyse the interaction of the lowest modes of the probe scalar fields resulting from the dimensional reduction (S-wave) with the fields in the JT theory. The JT$_\Ll$ description with $\Ll\neq 0$ is obtained by working with coordinates $\{r,\tau,z\}$ while the JT description ($\Ll=0$) is obtained by working with $\{r,t,z\}$; therefore the lowest Kaluza-Klein modes of the 3d probe scalar fields would be different. 
It would be interesting to analyse how the 2d scalar fields discussed in the last section for a fixed $\Ll$ give rise to the expected  OTOC behaviour of the scalar fields in 3d $i.e.$ the sawtooth like behaviour
\footnote{The JT theory is obtained by solving for the field strength in the 2d theory by assuming stationarity. Fluctuations about this stationary value of the field strength would give rise to a phase mode \cite{Moitra:2019bub} in the 2d theory. The interaction between the higher KK scalar modes and the phase mode would also have to be taken into account.}. 
Given that the left and right moving temperatures of the dual CFT$_2$ are $\frac{2\pi}{\beta_\pm}=\frac{2\pi}{\beta(1\pm \mu)}$, it isn't surprising that the temperature in the JT$_\Ll$ description is bounded by $\Ll<1$ to be $\frac{2\pi}{\beta(1-\mu\,\Ll)}<\frac{2\pi}{\beta_-}$. A  more complete picture of the dynamics of the JT$_\Ll$ description can be obtained by studying the dimensionally reduced 3d scalar fields and analysing their interactions with the fields in the JT$_\Ll$ theory, we leave this exercise for the near future.
\\\\
The formalism of the $\epsilon$-modes $-$or the ``reparametrization modes'', discussed in the last section to obtain the $\mathcal{O}(G_N)$ corrections to the 4pt functions in the JT theory have since been generalised to arbitrary $d$-dimensional CFTs in \cite{Haehl:2019eae}. These have a conformal dimension of -1 and can be seen as a tool to obtain the contribution of the stress-tensor conformal blocks to 4pt functions. The stress tensor and its shadow can be seen as descendants of these $\epsilon$-modes. Their 2pt functions in a $d$-dimensional CFT (without a temperature or chemical potential when $d>2$) can be used to obtain the stress-tensor conformal block contribution to a 4pt function \cite{Haehl:2019eae}, much the same way as used in section 5. The authors of \cite{Haehl:2019eae} also  find an action which reproduces the expected $\bra\epsilon\epsilon\ket$ correlator from general conformal invariance $i.e.$ an anomaly action for even $d$. Given the appearance of the $\epsilon$-modes in the near horizon analysis it seems interesting to speculate upon the bulk analogues of these modes in a holographic setting\footnote{Pedagogically the $\epsilon$-modes first were discovered in  the near horizon setting where the JT model was used to describe near extremal dynamics \cite{Jensen:2016pah,Maldacena:2016upp}. Following the renewed analysis of 3d gravity in AdS by Cotler and Jensen \cite{Cotler:2018zff} it was subsequently generalized in \cite{Haehl:2019eae} for any CFT$_d$. }. The $\epsilon$-modes discussed in \cite{Haehl:2019eae} were indeed described as degrees of freedom capturing the chaotic behaviour of the CFT. It is therefore not far fetched to speculate that the $\epsilon$-modes appearing the JT or the JT$_\Ll$ analysis are the dimensionally reduced near extremal analogues of the $\epsilon$ modes described in \cite{Haehl:2019eae} for a CFT$_d$. Holographically understanding the $\epsilon$-modes for AdS$_d$ black holes -along with the prescription of going near horizon and near extremal furnished in section 3, would provide a better understanding of the UV completion of the IR degrees of freedom we see in the JT analysis.
\\\\
There have been many works investigating whether gravity in $AdS_3$ can be considered a complete quantum theory in its own right \cite{Witten:2007kt,Maloney:2007ud,Cotler:2018zff,Maxfield:2020ale,Chandra:2022bqq}. The understanding gained from the analyses of \cite{Cotler:2018zff} is that the effective gravitational theory can be described as one of boundary reparametrizations provided one gives up modular invariance. Such a theory was shown to correctly contain the Virasoro contributions to the stress-tensor block in correlators of dual CFT \cite{Fitzpatrick:2015dlt,Fitzpatrick:2015zha}. 
The near extremal geometry has been known
to be sensitive to one of the left-right moving modes who’s
temperature approaches zero [37,50,51]. In the context of
the JT theory this was shown in [23,34] and in the context
of this paper it corresponds to the $\Ll=0$ case.
The full UV behaviour on the other hand is known to contain the dynamics of the Schwarzian action associated with both the left-right moving modes. 
This was also shown recently in context of bulk factorization for gravity in $AdS_3$ \cite{Mertens:2022ujr} and first in the context of computing Lyapunov index in generic BTZ \cite{Poojary:2018esz}. In the latter, the product of the 2 Schwarzian actions was obtained as the on-shell bulk action. In both cases the Schwarzian action/partition function for the left-right mode sees a temperature associated with the respective mode. Therefore the results of this paper can be viewed as accessing the different IR sectors of such a theory by choosing a canonical ensemble dependent on $\Ll$. The non-vanishing extremal temperature of one of the (left-right) modes of the CFT$_2$ can be achieved by taking the of extremal limit of $T_H/(1-\mu\Ll)$ along with  $\Ll\rightarrow\mu^{-1}$.     
\\\\
One can further generalize the new near horizon limits explored in this paper to the case of non-extremal geometries. Near horizon limits for non-extremal geometries have been explored in \cite{Godet:2021cdl} where the near horizon metric is obtained for arbitrary black hole geometries far from extremality. Here the authors show that the near horizon 2d dilaton theory is dual to the $\widehat{\rm CGHS}$ theory explored in \cite{Afshar:2019axx}. Although unlike the extremal and near extremal cases the near horizon metric in this case does not satisfy the Einstein's action satisfied by the parent black hole geometry, the near horizon theory does capture the constant charge thermodynamics of the parent geometry.  
\\\\
It would useful to understand the near horizon description presented in this paper for the case of higher dimensional near extremal rotating black holes, both in locally flat and $AdS$ spaces. Given that the bulk to boundary correlators are difficult to compute in all but the most symmetric of the $AdS$ spaces in dimensions $>3$ such a near horizon description is the best method for understanding the late time behaviours of the OTOC \cite{Blake:2021hjj}. It was also noted in \cite{Malvimat:2021itk,Malvimat:2022oue} that the presence of electric charge does not effect the blueshift suffered by a shockwave however it was recently noted in \cite{Horowitz:2022ptw} that charged shockwaves do bounce off after they have passed the outer horizon. This bounce produces a perceptible delay in the onset of scrambling which does not scale with the entropy of the black hole and is proportional to $\frac{\beta}{2\pi}\log(1-q\,Q)$ for AdS Reisnner-Nordstrom black holes where $q$ is the chemical potential and $Q$ is the charge per unit of the shockwave's energy. This bounce is also detectable in terms of the scalar field dynamics of the near horizon JT theory. It would be interesting to understand this near horizon perspective along with the higher dimensional generalization of the near horizon limits presented here in the case of the near extremal Kerr-Newman BHs in dimensions $>3$.          
\section*{Acknowledgements} The author thanks Daniel Grumiller for sharing his comments and insights on  this paper. The author is grateful to Ronak Soni for discussions and  comments on a previous version of this paper.  The author is supported by the Austrian Science Fund FWF $via$ the Lise Meitner project FWF M-2882 N.

\bibliographystyle{JHEP.bst}
\bibliography{bulk_syk_soft_modes.bib}
\end{document}